\documentclass[aps,prc,preprint,twocolumn,a4paper,10pt,superscriptaddress,floatfix,showpacs]{revtex4-1}

\usepackage{graphicx}
\usepackage{subcaption}
\usepackage{placeins}
\usepackage{amsmath}

\usepackage[utf8]{inputenc}

\newcommand{\hzdr}{\affiliation{Helmholtz-Zentrum Dresden-Rossendorf, Germany}}
\newcommand{\TUDD}{\affiliation{Technische Universit\"{a}t Dresden, Germany}}

\newcommand{\TUDa}{\affiliation{Technische Universit\"{a}t Darmstadt, Germany}}

\newcommand{\KVI} {\affiliation{KVI-CART, University of Groningen, The Netherlands}}

\newcommand{\EMMI}{\affiliation{GSI Helmholtzzentrum f\"ur Schwerionenforschung Darmstadt, Germany }}
\newcommand{\GSI} {\affiliation{GSI Helmholtzzentrum f\"ur Schwerionenforschung Darmstadt, Germany }}

\newcommand{\TUM} {\affiliation{Technische Universit\"{a}t M\"{u}nchen, Germany  }}

\newcommand{\CSICMadridSpain}{\affiliation{IEM CSIC, Madrid, Spain}}

\newcommand{\UniLisboa}{\affiliation{Centro de F\'{i}sica Nuclear da Universidade de Lisboa, Lisbon, Portugal}}

\newcommand{\UniSurrey}{\affiliation{University of Surrey, United Kingdom }}

\newcommand{\UniBirmingham}{\affiliation{School of Physics and Astronomy, University of Birmingham, United Kingdom }}

\begin{document}

\title{Coulomb dissociation of $^{20,21}$N
}

\author{Marko R\"oder}\hzdr \TUDD
\preprint{Accepted version, 01.06.2016}

\author{Tatsuya Adachi} \KVI

\author{Yulia Aksyutina} \EMMI

\author{Juan Alcantara}
\affiliation{ University of Santiago de Compostela, Spain }

\author{Sebastian Altstadt}
\affiliation{ Goethe University Frankfurt, Germany }

\author{Hector Alvarez-Pol}
\affiliation{ University of Santiago de Compostela, Spain }

\author{Nicholas Ashwood}\UniBirmingham 

\author{Leyla Atar}\GSI

\author{Thomas Aumann}
\TUDa
\GSI

\author{Vladimir Avdeichikov}
\affiliation{ Lund University, Sweden }

\author{M.~Barr}\UniBirmingham

% ...
\TUDD

\collaboration{R3B Collaboration}
\noaffiliation

\date{\today}

\begin{abstract}

Neutron-rich light nuclei and their reactions play an important role for the creation of chemical elements.
Here, data from a Coulomb dissociation experiment on $^{20,21}$N are reported.
Relativistic $^{20,21}$N ions impinged on a lead target and the Coulomb dissociation cross section was determined in a kinematically complete experiment.
Using the detailed balance theorem, the $^{19}\mathrm{N}(\mathrm{n},\gamma)^{20}\mathrm{N}$ and $^{20}\mathrm{N}(\mathrm{n},\gamma)^{21}\mathrm{N}$ excitation functions and thermonuclear reaction rates have been determined.
The $^{19}\mathrm{N}(\mathrm{n},\gamma)^{20}\mathrm{N}$ rate is up to a factor of 5 higher at $T<1$\,GK with respect to previous theoretical calculations, 
leading to a 10\,\% decrease
in the predicted fluorine abundance.

\end{abstract}


\maketitle 

\section{Introduction}
\label{sec:Introduction}

The astrophysical r-process (rapid neutron capture process) is an important process for the synthesis of heavy elements \cite{Burbidge1957}.
The path of the r-process involves many neutron-rich nuclei. 
One possible astrophysical site for the r-process are supernovae with a neutrino-driven wind scenario \cite{Takahashi_1994} where the neutrino wind dissociates all previously formed nuclei into protons, neutrons and alpha-particles \cite{Sasaqui_2005,Reifarth_2014a}. 
Nuclear reaction network calculations have shown that also light neutron-rich nuclei have an important impact on the final elemental abundance of the r-process nucleosynthesis \cite{Sasaqui_2005,Terasawa_2001}.
As the half-lives of nuclei close to the neutron drip line are very short, no target material can be fabricated.
Therefore, these nuclei have to be studied in beam, e.g. by exploiting the virtual gamma field of a lead target.
The astrophysically important neutron capture reaction may, then, be studied 
by time inversion applying the principle of detailed balance \cite{Baur1986}.

In the S393 experiment at the LAND/R3B setup (Large Area Neutron Detector; Reactions with Relativistic Radioactive Beams) at GSI Darmstadt, Germany, many neutron-rich nuclei were provided in a cocktail beam.
In this article, experimental results on the Coulomb dissociation cross sections of $^{20}$N and $^{21}$N are discussed.
Data on neutron-rich boron, carbon, and oxygen isotopes from the S393 experiment have been presented elsewhere \cite{Altstadt2014,Caesar2013,Heine16-arxiv,Thies16-PRC}.

\section{Experimental Setup}
\label{sec:ExperimentalSetup}

A primary $^{40}$Ar beam
with 490\,MeV/u kinetic energy
is guided onto a 4\,g/cm$^2$ thick Be target placed at the entrance of the FRS (FRagment Separator) \cite{Geissel_FRS_1992} producing a large variety of secondary ions in a cocktail beam.
Thereafter, the secondary ions pass a separation stage consisting of bending magnets and a fixed beam line.
As their beam trajectory is fixed by the magnetic rigidity $B\rho$, the secondary ions are separated with respect to their velocity according to their mass-to-charge ratio.

At the end of the FRS the secondary ions traverse a 3\,mm thick plastic scintillator (called S8) at a distance of 55\,m from the reaction target (mentioned later in detail). 
Additionally, the POS detector, a 1\,mm thick plastic scintillator at the entrance of the LAND/R3B cave, is placed 1.45\,m in front of the reaction target. 
Together, these detectors are used for time-of-flight measurements to identify the mass-to-charge ratio $A/Z$ of the secondary ions.

\begin{figure}
      \includegraphics[width = 0.48\textwidth]{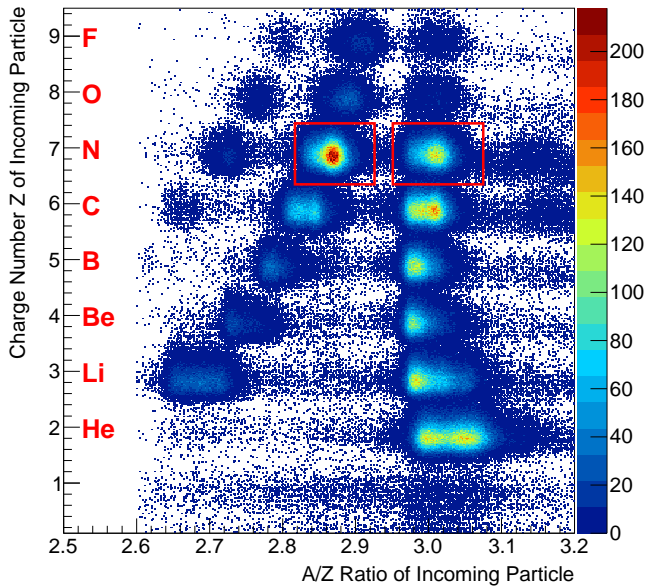}
      \caption{ Color online. 
        Incoming particle identification for the FRS setting used in this work. The charge number $Z$ is derived by energy loss measurements in the PSP and the $A/Z$ ratio by time-of-flight measurements between S8 and POS (55\,m distance).
        The red rectangles indicate a $3\sigma$-cut for $^{20}$N (left) and $^{21}$N (right). For further details, see text.
      }
      \label{fig:InPID}
\end{figure}

Besides that, a PSP detector (Position-Sensitive silicon Pin diode), placed 105\,cm upstream of the reaction target, is used to determine the charge of the secondary ions by an energy loss measurement $dE/dx$, which completes the identification of the particles impinging onto the LAND/R3B setup.
After a rough selection of the ion of interest, a Gaussian fit is applied to $A/Z$ and $dE/dx$ separately for each nucleus under study ($^{20}$N and $^{21}$N). 
All events within $3\sigma$ of the Gaussian fit are selected for further analysis (Figure \ref{fig:InPID}).
Contamination due to detector resolutions and cuts are studied in section \ref{sec:ErrorBudget}.

\begin{figure*}
    \includegraphics[width = 0.99\textwidth]{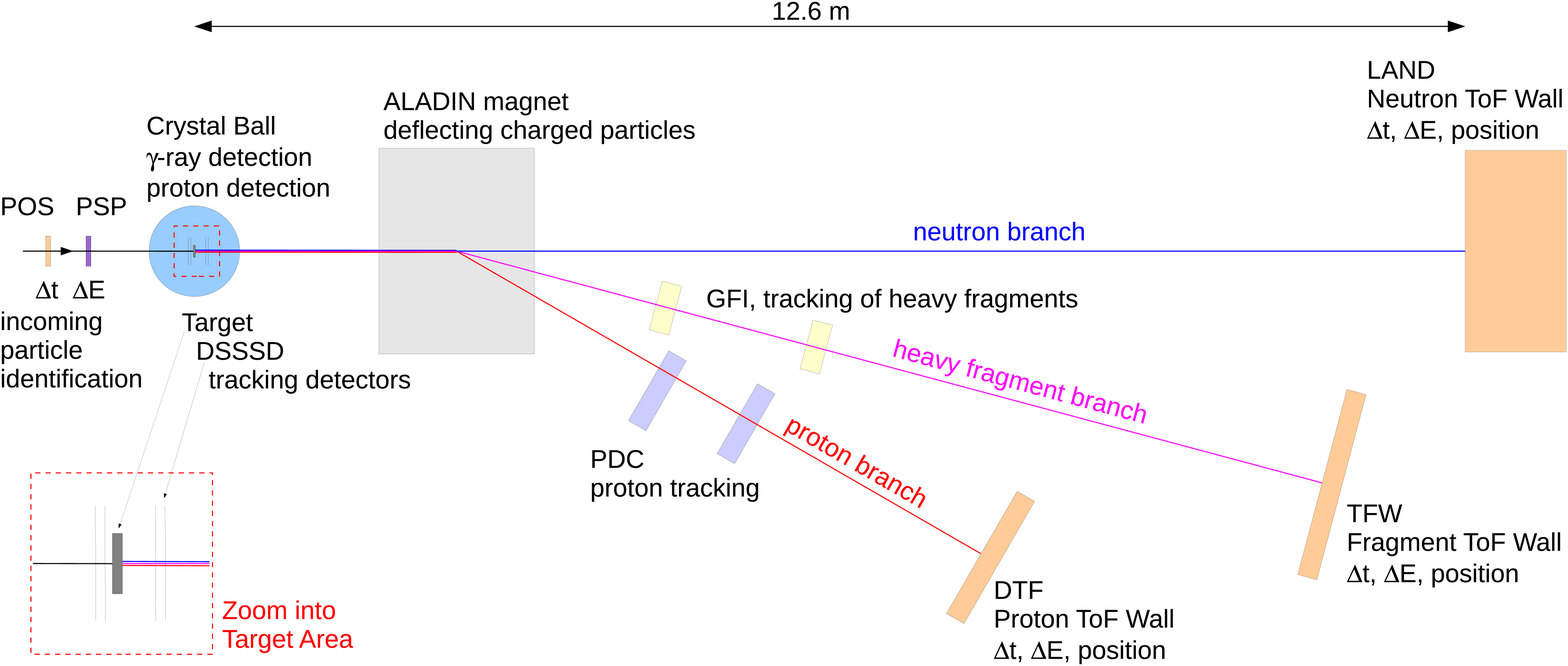}
    \caption{
      Sketch (top view) of the detector setup at the LAND/R3B cave. 
      Indicated are plastic scintillators (orange), the neutron flight path (blue line), the flight path of heavy ions (magenta line), and that of protons (red line).
Distances of the detectors are to scale, while the thicknesses of the thin detectors are not to scale. See text for details.}
    \label{fig:R3BSetup}
\end{figure*}

Furthermore, an active slit detector (called ROLU) consisting of four movable plastic scintillator sheets (5\,mm thickness), defining an empty rectangular window in the center, is utilized to define the accepted beam spot size.
Ions that do not traverse through the central window of ROLU but hit the active part of the detector are not recorded.

Moreover, four DSSSDs (Double Sided Silicon Strip Detector) \cite{Alcaraz2008_DSSSD} were placed in the beam line, two in front of and two behind the reaction target, to measure the track of the impinging particles and of the charged reaction products leaving the target.
A valid signal in all four DSSSDs is required for further analysis (see section \ref{sec:EventSelection}).
Additional four DSSSDs placed in a box arround the target to enable the detection of low energetic charged particles are not used for the present analysis.

Reaction targets with an area of 3$\times$3 cm$^2$ and a specific thickness $d$ ($d_\mathrm{Pb} = (0.176 \pm 0.004)$\,mm, $d_\mathrm{C} = (5.08 \pm 0.10)$\,mm) are mounted in a remotely controllable target wheel.
The lead target is used to study the Coulomb dissociation, 
while the data measured with the carbon target are used to subtract the nuclear contribution.
Measurements with no target are used to quantify the background contribution of the material in the beamline.

Around the reaction target, the Crystal Ball detector \cite{VMetag1983a}, consisting of 162 sodium iodine crystals arranged in a shell, is placed to detect $\gamma$ rays stemming from the deexcitation of excited states of the outgoing ions.
The proton detection capability of the Crystal Ball was not needed for the present analysis.

Downstream at 260\,cm from the reaction target, ALADIN (A LArge DIpole magNet) deflects charged particles according to their magnetic rigidity. 
At an angle of 15$^\circ$ from the nominal beam axis, two GFI detectors (Great FIbre detector) \cite{Cub1998_GFI,Mahata2009_GFI} with an active area of 50$\times$50\,cm$^2$
measure the x-position of the outgoing particles in order to identify the mass of the outgoing charged fragment.
The scintillating fibres with a cross section of 1$\times$1\,mm$^2$ are coupled to a position sensitive photomultiplier (PSPM) resulting in a spatial resolution of 1\,mm.
Finally, the charge, as well as the time-of-flight of the outgoing charged fragments are measured at the TFW (Time-of-Flight Wall), 
consisting of two crossing planes of plastic scintillator paddles with an active area of 189$\times$147\,cm$^2$.

The neutrons originating from Coulomb breakup reactions are unaffected by the magnetic field of ALADIN and impinge onto LAND (Large Area Neutron Detector)
\cite{TBlaich1992}, a 2\,m long, 2\,m wide and 1\,m thick device for detecting neutrons with kinetic energies between $T_\mathrm{n} = $ 100 and 1000\,MeV, placed at a distance of 12.6\,m from the reaction target.
LAND consists of 10 crossing planes of 20 paddles each. Each paddle (10\,cm thick) consists of consecutively 5\,mm iron converter (to convert the neutrons into detectable charged particles) and 5\,mm plastic scintillator sheets.
Further details can be found elsewhere \cite{TBlaich1992, KBoretzky2003}.

A sketch of the LAND/R3B setup as it was used for the S393 experiment is shown in Figure \ref{fig:R3BSetup}. 
Although not needed for the present analysis, the proton detection capabilities are shown for completeness.

\begin{figure*}
  \begin{subfigure}{0.49\textwidth}
     \includegraphics[width = 1\textwidth]{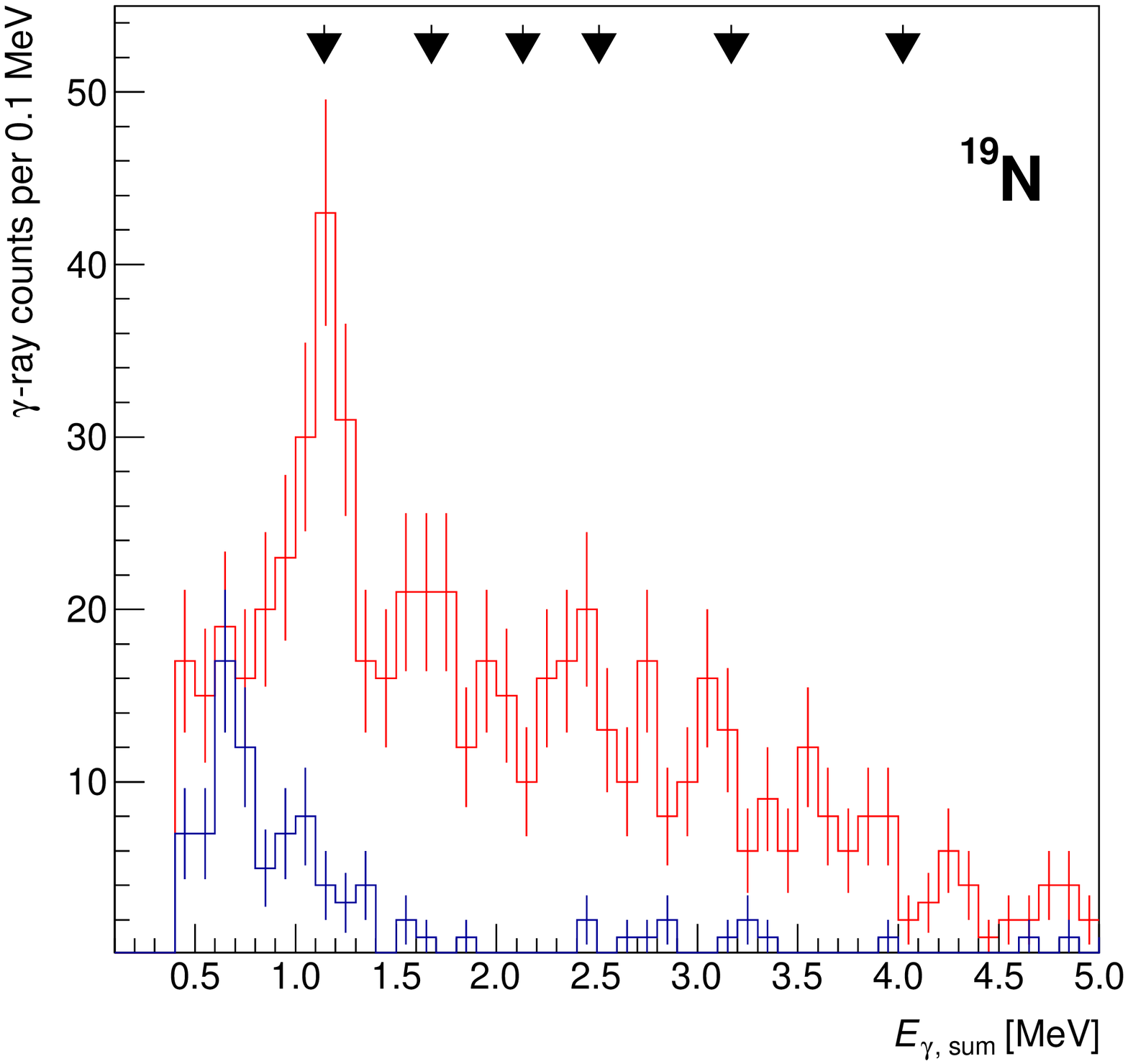} 
  \end{subfigure}
\hfill
  \begin{subfigure}{0.49\textwidth}
     \includegraphics[width = 1\textwidth]{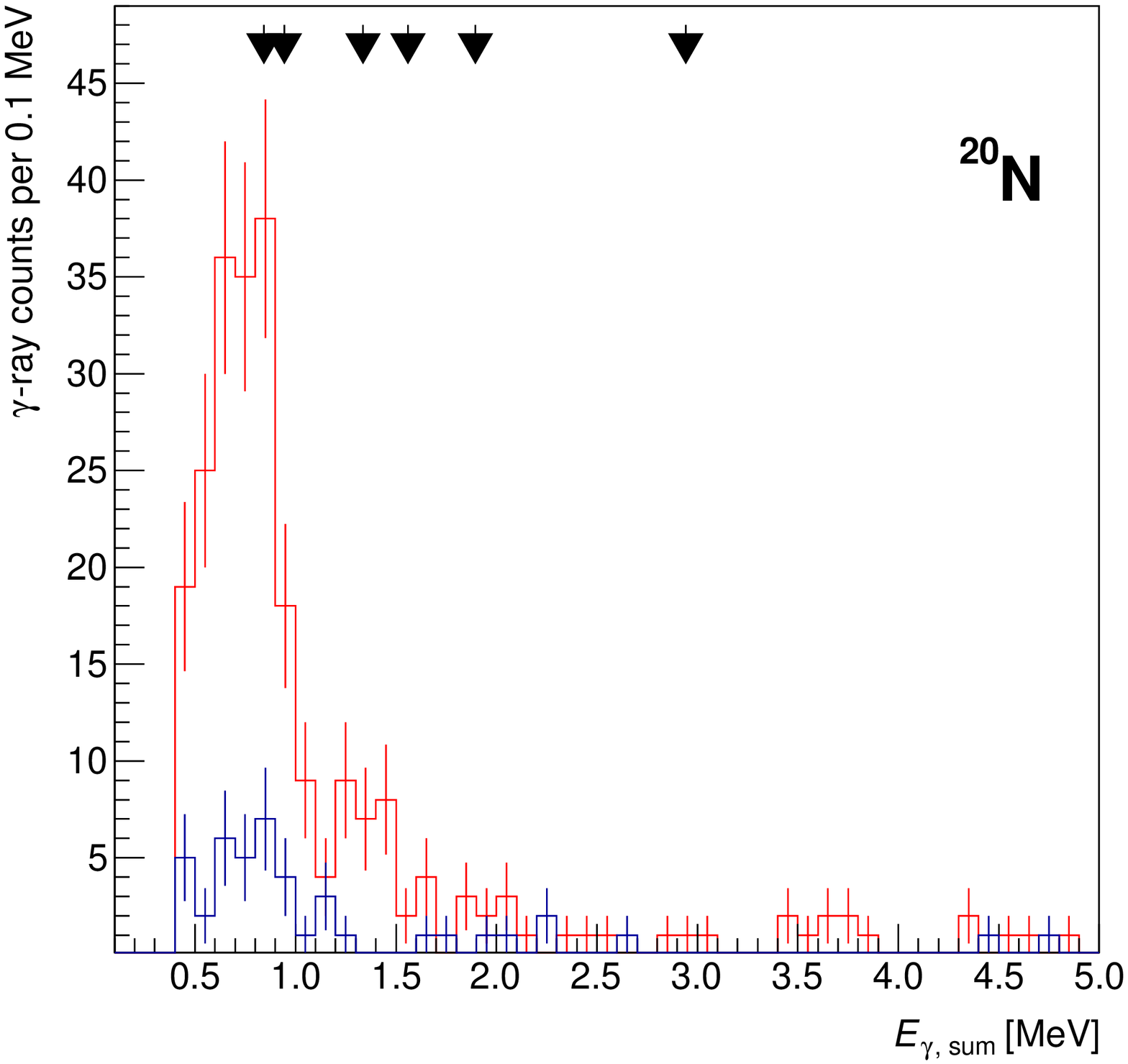}
  \end{subfigure}
  \caption{ Color online. 
Doppler corrected gamma sum spectra for outgoing
nuclei (red, solid line) after Coulomb breakup and a background estimation (blue, dashed line).
Left panel for outgoing $^{19}$N and right panel for $^{20}$N. 
       The black arrows indicate known excited states \cite{Sohler2008}.
       }
  \label{fig:GammaSumSpectra}

\end{figure*}

\section{Data Analysis}
\label{sec:DataAnalysis}

In this section, the analysis of the recorded events is discussed. The identification of the incoming beam was already described in the previous section.

\subsection{Selection of the Reaction Channel}
\label{sec:EventSelection}

After selecting the nuclei of interest during the incoming particle identification, we fix the charge number $Z$ of the outgoing heavy reaction fragment for the specific reaction channel ($Z$ = 7) by a cut on the energy loss in the TFW.
Then the deflection of the fragments in the magnetic field depends only on their mass number $A$ and their velocity.

In addition, the horizontal position measurement at the GFIs, used to identify the mass number of the fragment, is additionally correlated with the angle of emission and the interaction position at the target.
In order to convert the horizontal position measurements of the charged fragments at the GFIs into mass numbers, a mass reconstruction algorithm using the trajectories and time-of-flight information is used.

A valid reaction event is defined for further analysis when each of the following conditions are fulfilled.
Exactly one neutron is registered in LAND.
Late neutron events, i.e. from scattered neutrons, are rejected by a cut on the neutron velocity $v_\mathrm{n} > 20$\,cm/ns as this reflects the velocity of the incoming ion.
Exactly one outgoing $^{19}$N is detected in the fragment branch (both GFIs, and TFW).
In order to ensure the tracking of the incoming and outgoing particle, at least one hit in both planes of each of the four in-beam DSSSDs is required for data analysis.

\subsection{Study of Emitted $\gamma$ Rays}
\label{sec:GammaRays}

Transitions of the impinging nucleus into an excited state of the outgoing nucleus are identified by measuring the gamma energy released during the deexcitation of the state. 
As the particles are strongly forward boosted ($\beta_{^{20} \mathrm{N} } = 0.742$ and $\beta_{^{21} \mathrm{N} } = 0.726$, where $\beta$ is the velocity divided by the speed of light $c$), a correction for the Doppler shift is applied.
Furthermore, the segmentation of the Crystal Ball makes it necessary to take care of hits in several crystals by using an add\-back algorithm which works as follows.
Single crystals with an energy $E_{\gamma,\mathrm{crystal}}>0.3$\,MeV (to suppress background due to Bremsstrahlung) are grouped to clusters which are used to calculate the entire emitted gamma-energy during the reaction of one impinging particle.

In the left panel of Figure \ref{fig:GammaSumSpectra} (red, solid line) the summed gamma energy (sum of all clusters) for incoming $^{20}$N and outgoing $^{19}$N is shown.
A clear peak at $E_\mathrm{\gamma,sum} = 1.15$\,MeV
is visible corresponding to the first excited state of the outgoing $^{19}$N nuclei as previously measured by D.~Sohler~et~al. \cite{Sohler2008}.
The blue, dashed line indicates a background estimation by gating on unreacted outgoing $^{20}$N (otherwise equal cuts as discussed in section \ref{sec:EventSelection}). 
This shows that the peak at 1.15\,MeV is not caused by background events.
Furthermore, there is a high-energy tail with count rates half as large as that of the signal peak. These might correspond to higher lying states of $^{19}$N. But due to the limited statistics and energy resolution, they cannot be unambiguously identified.

The gamma sum spectrum originating from deexcitations of $^{20}$N is shown in the right panel of Figure \ref{fig:GammaSumSpectra}.
A dominant peak is visible at $E_\mathrm{\gamma,sum} = 0.85$\,MeV, again corresponding to the first excited state \cite{Sohler2008}.
Only a small additional structure arises at 1.3\,MeV and no significant structures are detected at $E_\mathrm{\gamma,sum}\ge1.5$\,MeV.
As D. Sohler et al. \cite{Sohler2008} reported several peaks in the close vicinity of 850\,keV, and due to the limited energy resolution of our gamma spectra, the observed peak cannot unambigiously be used to gate on the first excited state.
Therefore, a gate on transitions into 
the first plus the second excited state is given in the presented analysis.

\subsection{Coulomb Dissociation Cross Section}
\label{sec:CoulombDissociationCrossSection}

After applying the cuts for selecting the reaction channel, the reaction probabilities are 
obtained by dividing the number of reaction events (see section \ref{sec:EventSelection}) by the effective number of nuclei impinging onto the target.
The latter value is derived by counting all outgoing nitrogen isotopes that are detected at the fragment arm, including the unreacted $^{20}$N (resp. $^{21}$N) ions.
This automatically takes into account losses due to particles scattered out of the beamline, detector inefficiencies in the fragment arm, and applied cuts which together amount to 33\,\%.

The Coulomb dissociation cross section follows as
\begin{eqnarray}
 \sigma_\mathrm{CD} =   p_\mathrm{Pb}^\mathrm{react}  F_\mathrm{Pb}
                      - p_\mathrm{C}^\mathrm{react} \alpha F_\mathrm{C} 
                      - p_\mathrm{empty}^\mathrm{react}  \left ( F_\mathrm{Pb} - \alpha F_\mathrm{C} \right )                      
 \\
 \nonumber  \text{ with } F_\mathrm{target} = \frac {M_\mathrm{target}} {d_\mathrm{target} N_\mathrm{A}} ~\text{,}
\label{eq:CoulExCS}
\end{eqnarray}
where
   the indices Pb, C, and empty denote the specific targets (lead, carbon, and no target),
  $ p_\mathrm{target}^\mathrm{react}  $  is the probability of the impinging particle to react with the target,    
  $ M_\mathrm{target}  $  is the molar mass of the material of the target,                                                  
  $ d_\mathrm{target}  $  is the areal density of the target (measured by weighing),
  $ N_\mathrm{A}       $  is the Avogadro's number,                                                
  $ \alpha             $  is the nuclear scaling factor derived in the following paragraph.

The factor $\alpha$ is necessary to scale the nuclear contribution measured with the carbon target to the much larger lead nuclei. The black disk model is used to estimate
\begin{equation}
\alpha_\mathrm{^{20}N} = \frac{A_\mathrm{^{20}N} ^{1/3} + A_\mathrm{Pb} ^{1/3} } {A_\mathrm{^{20}N} ^{1/3} + A_\mathrm{C} ^{1/3}}  = 1.7 ~\text{.}
\label{eq:ScalingFactor}
\end{equation}
As $p_\mathrm{C}^\mathrm{react} < 0.1 \, p_\mathrm{Pb}^\mathrm{react}$, the choice of $\alpha$ has only a limited influence on $\sigma_\mathrm{CD}$.

\subsection{Excitation Energy}
\label{sec:ExcitationEnergy}

The
excitation energy $E^*$ of the reaction is extracted by the invariant mass method via
\begin{eqnarray}
 \label{eq:ExcitationEnergy}
  E^* =  && c^2 \sqrt{
            m_\mathrm{frag}^2 + m_\mathrm{n}^2 +
              E_x
         } 
         - m_\mathrm{proj}c^2 + E_\mathrm{\gamma,sum} 
 \\
 \nonumber \textrm{with}
 \\
 \nonumber  E_x = && 2\cdot \gamma_\mathrm{frag} \gamma_\mathrm{n} m_\mathrm{frag} m_\mathrm{n}
               ( 1-\beta_\mathrm{frag} \beta_\mathrm{n}  \cos{  \theta_\mathrm{frag, n}   } )  ~\text{,}
\end{eqnarray}
where  
       $m_\mathrm{proj}$ is the rest mass of the incoming nucleus,
       $m_\mathrm{frag}$ is the rest mass of the outgoing heavy reaction fragment,
       $m_\mathrm{n}$    is the rest mass of the outgoing neutron,
       $\beta_\mathrm{frag}$ is the velocity of the outgoing heavy reaction fragment,
       $\beta_\mathrm{n}$    is the velocity of the outgoing neutron,
       $\gamma_\mathrm{frag} = (1-\beta^2_\mathrm{frag})^{-1/2}$,
       $\gamma_\mathrm{n} =  (1-\beta^2_\mathrm{n})^{-1/2} $,
       $\theta_\mathrm{frag, n}$ is the angle between the outgoing heavy reaction fragment and the neutron, while
       $E_\mathrm{\gamma,sum}$   is the energy of all gamma quanta emitted during the reaction. Here, due to the high photo-peak efficiency ($\sim$70\%) of the crystal ball at the relatively low $\gamma$-ray energies ($\sim$1\,MeV), $E_\mathrm{\gamma,sum}$ is simply taken as the sum of the detected $\gamma$-ray energies. This approximation entails a slight downward shift of the excitation function for Coulomb dissociation into excited states of the product nucleus, which is not significant here, because those states contribute only negligibly to the reaction rate at the effective astrophysical temperatures, see below.

The mass of the incoming $^{20}$N (resp. $^{21}$N) and the outgoing $^{19}$N (resp. $^{20}$N) were taken from 
the AME2003 mass evaluation \cite{GAudi2003}.
Differences to the updated mass evaluation AME2012 \cite{AME2012_Wang} amount to only 10\,keV, much less than the experimental resolution.

An energy-dependent correction for the neutron detection efficiency must be applied to the Coulomb dissociation cross section.
The total neutron efficiency of LAND is based on a simulation which includes the acceptance due to the kinematics of the specific reaction, the acceptance due to deactivated/broken paddles and the nominal energy-dependent neutron efficiency of the detector measured in an earlier experiment \cite{KBoretzky2003}.

For kinetic energies in the center of mass of $T_\mathrm{n}^\mathrm{c.m.} < 5$\,MeV, the neutrons are strongly forward boosted,
so that all neutrons hit the active area of LAND.
The efficiency drops dramatically for $T_\mathrm{n}^\mathrm{c.m.} > 5$\,MeV as the transverse momentum component becomes more dominant and, thus, many neutrons miss the active area of LAND.

\begin{figure}
    \includegraphics[width = 0.48\textwidth]{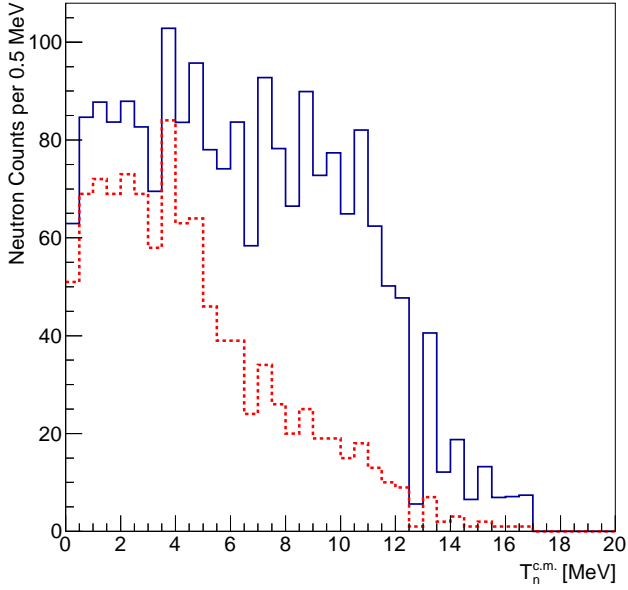}
    \caption{ Color online. 
      One-neutron emission spectrum of $^{20}$N impinging onto a Pb target measured with LAND (red, dashed line) and corrected for the LAND efficiency (blue, solid line). The kinetic energy of the neutron $T_\mathrm{n}^\mathrm{c.m.}$ is given in the center of mass frame.
    }
    \label{fig:NeutronEnergy}
\end{figure}

The kinetic energy of the neutrons in the center of mass system for the $^{20}\mathrm{N}(\gamma,\mathrm{n})^{19}\mathrm{N}$ reaction is depicted in Figure \ref{fig:NeutronEnergy}.

\begin{table*}
  \caption{
    Summary of the Coulomb dissociation cross sections of $^{20}$N and $^{21}$N integrated over $E^* \in [0,20]$\,MeV for
    the total reaction,
    transitions into any excited state of the outgoing nucleus,
    transitions into the ground state of  the outgoing nucleus, and
    transitions into the first excited state of the outgoing nucleus (only for $^{20}\mathrm{N}(\gamma^*,\mathrm{n})^{19}\mathrm{N}$ reaction). 
    Only the statistical uncertainties are given here, as they are much larger than the systematic ones (see text for details).
    }
  \label{tab:CoulexSummaryErrors_20N}
    \begin{tabular}{ l l l l r l l }
    \hline
    \hline
    $\sigma_\mathrm{CD}(^{20}$N, total)                           & = & (90 & $\pm $ & $12^\mathrm{stat}$& ) & mb  \\
    $\sigma_\mathrm{CD}(^{20}$N, all excited states)              & = & (76 & $\pm $ & $10^\mathrm{stat}$& ) & mb  \\
    $\sigma_\mathrm{CD}(^{20}$N, ground state)                    & = & (15 & $\pm $ & $16^\mathrm{stat}$& ) & mb  \\
    $\sigma_\mathrm{CD}(^{20}$N, 1st exc. state)                  & = & (36 & $\pm $ & $ 6^\mathrm{stat}$& ) & mb  \\
    \hline                                                                                      
    \hline                                                                                      
    $\sigma_\mathrm{CD}(^{21}$N, total)                           & = & (75 & $\pm $ & $13^\mathrm{stat}$& ) & mb  \\
    $\sigma_\mathrm{CD}(^{21}$N, all excited states)              & = & (44 & $\pm $ & $ 9^\mathrm{stat}$& ) & mb  \\
    $\sigma_\mathrm{CD}(^{21}$N, ground state)                    & = & (31 & $\pm $ & $16^\mathrm{stat}$& ) & mb  \\
    $\sigma_\mathrm{CD}(^{21}$N, 1st + 2nd exc. state)            & = & (47 & $\pm $ & $ 8^\mathrm{stat}$& ) & mb  \\
    \hline
    \hline
    \end{tabular}
\end{table*}  

%
%
%
%

\subsection{Differential Coulomb Dissociation Cross Section}

The differential reaction cross section of $^{20}$N impinging onto the lead target as a function of the excitation energy $E^{*}$ is extracted from the yield formula

\begin{eqnarray}
\nonumber
\frac { d\sigma_\mathrm{Pb} } {dE^*}
 = 
\left(
        \int\limits_{E^*}^{E^*+dE}
          \frac {d\sigma_\mathrm{Pb} ( \tilde E ) } { d \tilde E } d \tilde E  
\right)
        \frac {1} {\Delta E^*}
\\
 =  
  \frac { N(\mathrm{^{19}N}) (E^* ) } { N(\mathrm{^{20}N}) \eta_\mathrm{LAND} } F_\mathrm{ Pb } 
  \frac { 1 } {\Delta E^* } ~\text{,}
\end{eqnarray}
where 
$N(\mathrm{^{19}N})$ is the number of outgoing $^{19}$N nuclei,
$N(\mathrm{^{20}N})$ is the number of incoming $^{20}$N nuclei,
and
$\eta_\mathrm{LAND}$ is the one-neutron detection efficiency of LAND.

The differential Coulomb dissociation cross section as a function of the excitation energy (Eq. \ref{eq:ExcitationEnergy}), then, is given by

\begin{eqnarray}
\frac { d\sigma_\mathrm{CD} } {dE^*} 
 =  
  \,
  \nonumber
   &&  \left[ \, p_\mathrm{Pb}(E^*)        F_\mathrm{Pb} - \right.
\\
   && \, \left. \, p_\mathrm{C}(E^*)  \alpha F_\mathrm{C}  - \right.
   \nonumber
\\
   && \, \left. \, p_\mathrm{empty}(E^*)
            F_\mathrm{empty }
   \, \right] \,
   \frac{1} {\Delta E^*} ~\text{.}
\end{eqnarray}

The total differential Coulomb dissociation cross section (including transitions into the ground state and any excited state) as a function of the excitation energy for impinging $^{20}$N is depicted in Figure \ref{fig:ExcitationEnergySpectrum_total_CDCS_20N} (black line). 

In order to separate transitions of 
$^{20}\mathrm{N}(\gamma^*,\mathrm{n})$ into the first excited state of $^{19}$N at 1141\,keV, a cut on 
$0.70\,\mathrm{MeV}<E_\mathrm{\gamma,sum}<1.40$\,MeV
is performed.
Furthermore, the resulting $E^{*}$ spectrum is corrected for the photo peak efficiency of the Crystal Ball $\epsilon (E_\mathrm{\gamma,sum} = 1.15\textrm{\,MeV}) = 0.67$ which is based on simulations taking into account the kinematic boost. The result is shown in Figure \ref{fig:ExcitationEnergySpectrum_total_CDCS_20N} (red line).

Transitions into the ground state of $^{19}$N are explored by subtracting transitions into any excited state (with a gate on $E_\mathrm{\gamma,sum}>0.7$\,MeV) from the total excitation energy spectrum (compare Figure \ref{fig:ExcitationEnergySpectrum_total_CDCS_20N}).
Here, the excited state data are corrected for the total $\gamma$ efficiency of 0.85, which is constant within 3\% for $E_\mathrm{\gamma,sum}$ = 1-5\,MeV.

The total differential Coulomb dissociation cross section as a function of the excitation energy for impinging $^{21}$N is shown in Figure \ref{fig:ExcitationEnergySpectrum_total_CDCS_21N} (black line).
Transitions of $^{21}\mathrm{N}(\gamma^*,\mathrm{n})$ into the first plus second excited state of $^{20}\mathrm{N}$ are deduced by gating on $0.40\,\mathrm{MeV}<E_\mathrm{\gamma,sum}<1.00$\,MeV and correcting for the photo peak efficiency of $\epsilon (E_\mathrm{\gamma,sum} = 0.90\textrm{\,MeV})$.

In order to derive the transitions into the ground state of $^{20}\mathrm{N}$ (Fig. \ref{fig:ExcitationEnergySpectrum_total_CDCS_21N}, magenta line), the $E^{*}$ spectrum of the transitions of $^{21}\mathrm{N}(\gamma^*,\mathrm{n})$ into any excited state of $^{20}\mathrm{N}$ with $E_\mathrm{\gamma,sum}>0.40$\,MeV is corrected with the total $\gamma$ efficiency and subtracted from the total reaction.

\subsection{Error Budget}
\label{sec:ErrorBudget}

Systematic uncertainties from the identification of the incoming particles are derived by varying the cuts on $A/Z$ and $dE/dx$ of the incoming particle from $2\sigma$ to $3\sigma$ and calculating the difference. As these uncertainties are very small (3\,\%) in contrast to the statistical uncertainties, these will be neglected in the further analysis.

Moreover, the single neutron detection efficiency of LAND is known with an uncertainty of 6\,\%.
The Crystal Ball efficiency is determined with an uncertainty of 6\,\%. 

The uncertainty of the measurement of the areal density of the target amounts to 2\,\% and, thus, will be neglected in the further analysis.

The results of the Coulomb dissociation cross sections and the corresponding uncertainties
are summarized in Table \ref{tab:CoulexSummaryErrors_20N}.

\section{Results}
\label{sec:Results}
Most of the Coulomb dissociating $^{20}$N nuclei populate excited states of $^{19}$N (Table \ref{tab:CoulexSummaryErrors_20N}): 
40\,\% of the total Coulomb dissociation cross section is caused by transitions into the first excited state of $^{19}$N at 1141\,keV, 44\% to higher excited states..
Only 17\,\% of the reactions are caused by transitions into the ground state which is compatible with zero within the statistical uncertainties.

\begin{figure}
    \includegraphics[width = 0.47\textwidth]{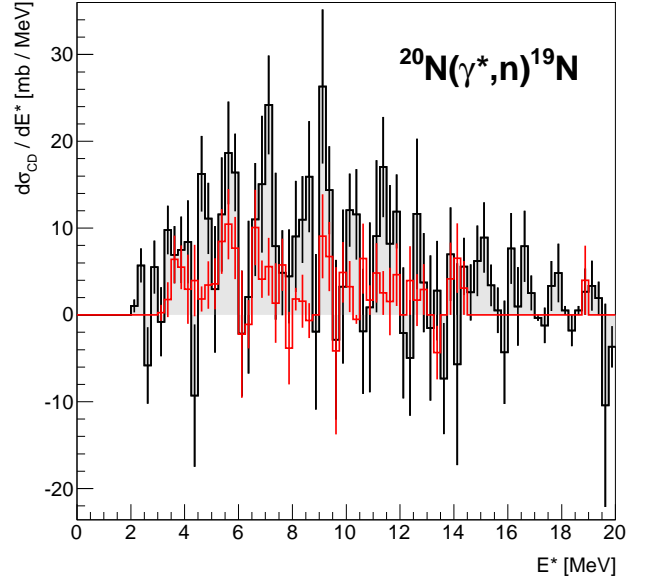}
    \caption{ Color online. 
      Coulomb dissociation cross section of $^{20}$N as a function of the excitation energy for the total reaction (black line) and transitions into the first excited state of $^{19}$N at 1141\,keV (red line).
      The error bars reflect statistical uncertainties only.
    }
    \label{fig:ExcitationEnergySpectrum_total_CDCS_20N}
 \end{figure}

The total excitation energy spectrum of the Coulomb dissociation of $^{20}$N is shown in Figure \ref{fig:ExcitationEnergySpectrum_total_CDCS_20N} (black line).
While there are no entries for energies from 0 to 2\,MeV, the spectrum notably increases at 2\,MeV.
The one-neutron separation threshold of $^{20}$N  is $S_\mathrm{1n}(^{20}\mathrm{N}) = 2.16$\,MeV.

Some structures appear at 5.5\,MeV, 7.0\,MeV, 9.0\,MeV, 10.2\,MeV and 11.5\,MeV.  
For $E^* > 17$\,MeV, the spectrum drops to values consistent with zero within the statistical errors as the number of virtual photons at 18\,MeV drops to 15\,\% of that at 3\,MeV.

Transitions of $^{20}$N into the first excited state of $^{19}$N at 1143\,keV are depicted in Figure \ref{fig:ExcitationEnergySpectrum_total_CDCS_20N} (red line).
Here, the spectrum rises significantly above zero for energies $E^{*} = 3.5$\,MeV which reflects the sum of $S_\mathrm{1n}(^{20}\mathrm{N})$ and the energy of the first excited state. 
Some structure is visible for energies between 3.5\,MeV and 14\,MeV.

A summary of the cross sections for the Coulomb dissociation of $^{21}$N is presented in Table \ref{tab:CoulexSummaryErrors_20N}.
Here, 59\,\% of the Coulomb dissociating $^{21}$N transit into excited states of $^{20}$N (less than for the $^{20}\mathrm{N}(\gamma^*,\mathrm{n})^{19}\mathrm{N}$ reaction).
41\,\% of the reaction pass into the ground state of $^{20}$N (higher than for $^{20}\mathrm{N}(\gamma^*,\mathrm{n})^{19}\mathrm{N}$).

\begin{figure}
    \includegraphics[width = 0.47\textwidth]{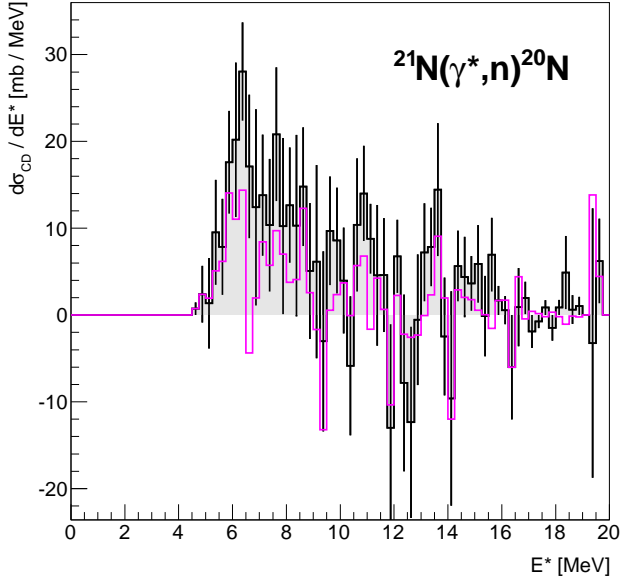}
    \caption{ Color online. 
      Coulomb dissociation cross section of $^{21}$N as a function of the excitation energy for the total reaction (black line) and transitions into the ground state of $^{20}$N (magenta line). 
      The error bars reflect statistical uncertainties only.
      For the ground state transitions, the error bars were omitted for reasons of clarity as they are comparable to the ones of the total reaction.
    }
    \label{fig:ExcitationEnergySpectrum_total_CDCS_21N}
\end{figure}

The total energy-dependent Coulomb dissociation cross section of $^{21}$N is shown in Figure \ref{fig:ExcitationEnergySpectrum_total_CDCS_21N} (black line).
The spectrum does not show any entry between 0\,MeV and 4.5\,MeV.
The one-neutron separation threshold of $^{21}$N is $S_\mathrm{1n}(^{21}\mathrm{N}) = 4.60$\,MeV.
The spectrum rises significantly above zero at $E^{*} = 5.25$\,MeV.
Subsequently, there is a structure between 5\,MeV and 14\,MeV.
Beyond 14\,MeV the spectrum drops to values consistent with zero.

The energy-dependent Coulomb dissociation cross section of $^{21}$N, passing into the ground state of $^{20}$N, is depicted in Figure \ref{fig:ExcitationEnergySpectrum_total_CDCS_21N} (magenta line).
The spectrum rises significantly above zero for energies between 5\,MeV and 9\,MeV with a plateau-like structure.
No significant peaks can be identified due to the large statistical uncertainties.

\section{Photo Absorption and Neutron Capture Cross Section}
\label{sec:PhotoAbsorption}

The differential photo absorption cross section is calculated via the virtual photon theorem \cite{Baur1986} 
\begin{equation}
 \frac{ d\sigma_\mathrm{CD} }{ dE_{\gamma}} = \frac{1}{ E_{\gamma} } n_\mathrm{E1}(E_{\gamma}) \sigma^\mathrm{photo}_\mathrm{E1} ~\text{.}
 \label{eq:Photoabsorption_CS}
\end{equation}
Here, $\frac{ d\sigma_\mathrm{CD} }{ dE_{\gamma}}$ is the differential Coulomb dissociation cross section as a function of the excitation energy (where $E_{\gamma} \equiv E^*$), calculated as in the previous section,
$n_\mathrm{E1}(E_{\gamma})$ is the number of virtual photons and
$\sigma^\mathrm{photo}_\mathrm{E1}$ is the photo absorption cross section for multipolarity E1 as a function of the excitation energy.

As the cross sections of higher multipolarities are found to be three orders of magnitude lower than for E1 \cite{Typel_HigherOrderEffectsElMagDissociation_2001}, higher multipolarities are neglected.

Then, 
the photo absorption cross section is calculated from
\begin{equation}
  \sigma_{\gamma,\mathrm{n}} \equiv \sigma^\mathrm{photo}_\mathrm{E1} = \frac{ d\sigma_\mathrm{CD} }{ dE^*} \frac{1}{ n_\mathrm{E1}(E^*) } E^*  ~\text{.}
 \label{eq:Photoabsorption_CS_3}
\end{equation}

The virtual photon spectrum for the E1 multipolarity is derived as described in detail in \cite{Baur1986} 
\begin{eqnarray}
 n_\mathrm{E1}(E^*) = \frac{2}{\pi} Z^2_\mathrm{T} e^2 
 \alpha \left(\frac{c}{v}\right)^2 \left( \xi K_0(\xi) K_1(\xi) - L(\xi) \right)
 \\
 \nonumber \text{ with } L(\xi) = \frac{v^2 \xi^2  }{2 c^2} \left[ K^2_1(\xi) -  K^2_0(\xi) \right] ~\text{,}
  \label{eq:VirtualPhotonSpectrumE1}
\end{eqnarray}
where $K_{i}$ are the modified Bessel functions of the order $i$, $\alpha$ the fine-structure constant, and $\xi$ the adiabaticity parameter 
which reads
 $\xi = { E^* b } / ({\hbar \gamma \beta c}) $
with $b$ as the impact parameter.

Utilizing the fact that nuclear reactions are invariant under time reversal, the neutron capture cross section is determined
via the theorem of detailed balance \cite{Baur1986}
\begin{equation}
 \sigma_\mathrm{\mathrm{n},\gamma} = \frac{ 2(2I_\mathrm{A}+1) }  { (2I_\mathrm{B}+1) ( 2I_\mathrm{n} +1 ) } \frac{k^2_{\gamma} } {k^2_\mathrm{c.m.} } \sigma_\mathrm{\gamma,\mathrm{n}} ~\text{,}
 \label{eq:NeutronCapture}
\end{equation}

\noindent where $\sigma_{\gamma,\mathrm{n}}$ is the photo absorption cross section (compare Eq. \ref{eq:Photoabsorption_CS_3}),
$k_{\gamma} = E^* /(\hbar c) $, 
$k^2_\mathrm{c.m.}   = 2\mu(E^* - Q) / \hbar^2 $ with $Q = S_\mathrm{1n} + E_\mathrm{\gamma,sum}$
and $\mu$ is the reduced mass of the system of the outgoing fragment plus neutron (e.g. for the system $^{19}$N+n,
$\mu = (M_\mathrm{^{19}N} \cdot M_\mathrm{n} ) / ( M_\mathrm{^{19}N} + M_\mathrm{n} ) $).
$I_{A,B}$ are the spins of the incoming and outgoing particle. 
With $^{20}$N as incoming nucleus, 
index A denotes $^{20}$N with 
$I_\mathrm{A}$ = 2 for the ground state, 
index B represents the outgoing $^{19}$N with $I_\mathrm{B}$ = 1/2 for the ground state 
and $I_\mathrm{B}$ = 3/2 for the first excited state, 
and index n indicates the neutron with $I_\mathrm{n}$ = 1/2.
With $^{21}$N as incoming beam, $I_\mathrm{A}$ = 2 for the ground state, 
index B represents the outgoing $^{20}$N with $I_\mathrm{B}$ = 2 for the ground state 
and $I_\mathrm{B}$ = 3 for the first excited state.

Integrating over 
energies $E_\mathrm{cm} \in [0,15]$\,MeV (where $E_\mathrm{cm} = E^* - Q$), 
the neutron capture cross section of 
the ground state of $^{19}$N amounts to $\sigma_\mathrm{n,\gamma}(^{19}\mathrm{N,g.s.}) = (0.003 \pm 0.010^\mathrm{stat})$\,mb.
As the statistical uncertainty is large, only an upper limit with 90\,\% confidence level is given
$\sigma_\mathrm{n,\gamma}(^{19}\mathrm{N,g.s.}) \le 0.016$\,mb. 
The neutron capture cross section of 
the first excited state of $^{19}$N amounts to \linebreak
$\sigma_\mathrm{n,\gamma}(^{19}\mathrm{N,1st}) = (0.0057 \pm 0.0014^\mathrm{stat})$\,mb.
Similarly, the neutron capture cross section of the ground state of $^{20}$N amounts to
$\sigma_\mathrm{n,\gamma}(^{20}\mathrm{N,g.s.}) \le 0.0091$\,mb.

The first and the second excited state of $^{20}$N cannot be clearly separated due to the low resolution of our gamma spectra.
For further analysis of the stellar reaction rate and its implementation into a reaction network we provide
the neutron capture cross section of 
the first and second excited state of $^{20}$N \linebreak
$\sigma_\mathrm{n,\gamma}(^{20}\mathrm{N,1st+2nd}) = (0.0041 \pm 0.0010^\mathrm{stat})$\,mb.

A summary of the derived photo absorption and neutron capture cross sections is given in Table \ref{tab:PhotoAbsorptionNeuCapCS}.

\begin{table}
  \caption{
    Photo absorption and neutron capture cross section of the nitrogen isotopes under study. The errors reflect statistical uncertainties only. Upper limits are given with 90\,\% confidence level.
  }
  \label{tab:PhotoAbsorptionNeuCapCS}
  \begin{ruledtabular}
    \begin{tabular}{  l l l l l l l l  }
    $\sigma_\mathrm{\gamma,n}(^{20}$N), 
                                     & total                & =     & (1.15    & $\pm$ & 0.27   & ) & mb   \\
                                     & ground state         & $\le$ & \ 0.62   &       &        &   & mb   \\
                                     & 1st excited state    & =     & (0.51    & $\pm$ & 0.12   & ) & mb   \\
    \hline
    $\sigma_\mathrm{\gamma,n}(^{21}$N),
                                     & total                & =     & (0.93    & $\pm$ & 0.30   & ) & mb   \\
                                     & ground state         & $\le$ & \ 0.74   &       &        &   & mb   \\
                                     & 1st + 2nd exc. state & =     & (0.69    & $\pm$ & 0.16   & ) & mb   \\
    \hline
    \hline
    $\sigma_\mathrm{n,\gamma}(^{19}$N), 
                                     & ground state          & $\le$ & \ 0.016  &       &        &   & mb   \\
                                     & 1st excited state)    & =     & (0.0057  & $\pm$ & 0.0014 & ) & mb   \\
    \hline
    $\sigma_\mathrm{n,\gamma}(^{20}$N), 
                                     & ground state          & $\le$ & \ 0.0091 &       &        &    & mb   \\
                                     & 1st + 2nd exc. state) & =     & (0.0041  & $\pm$ & 0.0010 & )  & mb   \\
   \end{tabular}
 \end{ruledtabular}
\end{table}

\section{Astrophysical Reaction Rate}
\label{sec:AstrophysicalReactionRate}

From the derived neutron capture cross section, 
the Maxwellian averaged reaction rate \cite{Angulo1999_ReactionRates} is calculated
\begin{equation}
N_\mathrm{A} \langle \sigma v \rangle  = N_\mathrm{A} \frac{ (8/\pi)^{1/2} }{ \mu^{1/2} (k_\mathrm{B} T )^{3/2} } \int\limits_{0}^{\infty} \sigma_\mathrm{n,\gamma} E \exp \left( - \frac{ E }{ k_\mathrm{B} T } \right) dE ~\text{,}
\label{eq:MaxAveReactRate}
\end{equation}
where $N_\mathrm{A}$ is the Avogadro's number, $\mu$ the reduced mass of the system under study (e.g. \mbox{$^{19}$N + n}), $k_\mathrm{B}$ the Boltzmann constant, and $T$ the temperature of the stellar environment (assumed to be in thermal equilibrium with the nuclei under study).

Due to the large, predominantly statistical uncertainty of the neutron capture cross section of the ground state of $^{19}$N and $^{20}$N, we applied a randomization technique.
For each energy bin in the measured neutron capture histogram, a random Gaussian distribution is built where the bin content acts as mean value and the statistical uncertainty of the bin acts as sigma of the Gaussian distribution.
Then, the randomly distributed neutron capture cross section is used to calculate the Maxwellian averaged reaction rate for each temperature (compare Eq. \ref{eq:MaxAveReactRate}), forcing the Maxwellian averaged reaction rates to be positive definite. This procedure is repeated 1000 times, thus, for each temperature we generate 1000 times a Maxwellian averaged reaction rate. Then, the mean value of the resulting distribution is used as a value in the randomized Maxwellian averaged reaction rate as a function of temperature while the root mean square (rms) acts as statistical uncertainty.

In order to judge how much the excited states contribute to the total reaction rate, 
the
population of the first excited state of $^{19}$N relative to the ground state is assumed to be \cite{Rolfs1988}
\begin{equation}
P(E_\mathrm{R}) = \exp \left( - \frac{ E_\mathrm{R} } { k_\mathrm{B} T } \right)  \frac { 2I_\mathrm{R} + 1 } { 2I_0 + 1 } ~\text{,}
\label{eq:Population1st19N5GK}
\end{equation}
where $E_\mathrm{R}$ is the energy of the excited state, $I_\mathrm{R}$ the spin of the resonant state and $I_0$ the spin of the ground state.

In order to calculate the stellar reaction rate $R$, we combine the neutron capture reaction rate of the ground state $R_0$ and the neutron capture reaction rate of the first excited state $R_1$ with the population probability (Eq. \ref{eq:Population1st19N5GK})  \cite{Sallaska_2013_StellarReactionRate}:
\begin{equation}
R = \frac { g_\mathrm{0}  R_\mathrm{0} + g_\mathrm{1}  R_\mathrm{1}  \exp\left( - \frac {E_\mathrm{1}} {k_\mathrm{B} T} \right) } 
    { g_\mathrm{0} + g_\mathrm{1} \exp \left( - \frac{E_\mathrm{1}}{k_\mathrm{B} T} \right) }
\label{eq:StellarReactionRate}
\end{equation}
with $g_\mathrm{0} = 2I_\mathrm{0} + 1$.

It should be noted that the dissociation cross sections were measured for nuclei in their ground state. Therefore, the cross section and reaction rate determined using the detailed balance theorem only constrain the contribution of the ground state. As a result, the contribution of low-energy first excited states, which are located at 843 and 1177 keV for $^{20,21}$N respectively \cite{Sohler2008}, to the total (n,$\gamma$) reaction rate is not constrained by the present measurement. In previous studies, the inclusion of the capture to thermally populated excited states has led to significant enhancement of the total cross section \cite{Heine16-arxiv}.

Then, the $^{19}\mathrm{N}(\mathrm{n},\gamma)^{20}\mathrm{N}$ reaction rates as a function of the temperature are compared with reaction rates included in reaction network codes which are mainly based on one theoretical work \cite{Rauscher1994} considering just direct capture and giving a linear approximation \mbox{$N_A \langle \sigma v \rangle ^\mathrm{theory} = 1.54 \cdot 10^3 \cdot T $}.

\begin{figure}
  \includegraphics[width = 0.47\textwidth,clip]{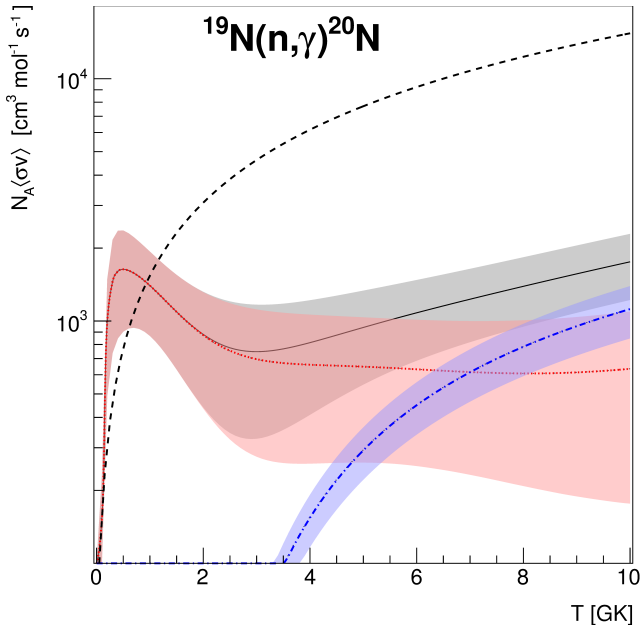}
  \caption{ 
      Color online. 
      Stellar reaction rate for $^{19}\mathrm{N}(\mathrm{n},\gamma)^{20}\mathrm{N}$ (black, solid line).
      The red, dotted line denotes contributions from the neutron capture of the ground state of $^{19}$N
      and the blue, dotted-dashed line denotes contributions from the neutron capture of the first excited state of $^{19}$N.
      The black, dashed line denotes a theoretical curve given by \cite{Rauscher1994}.
  }
  \label{fig:ReactionRate19N-ng-20N}
\end{figure}

\begin{figure}
  \includegraphics[trim = 0 0 0 0,clip,width = 0.47\textwidth]{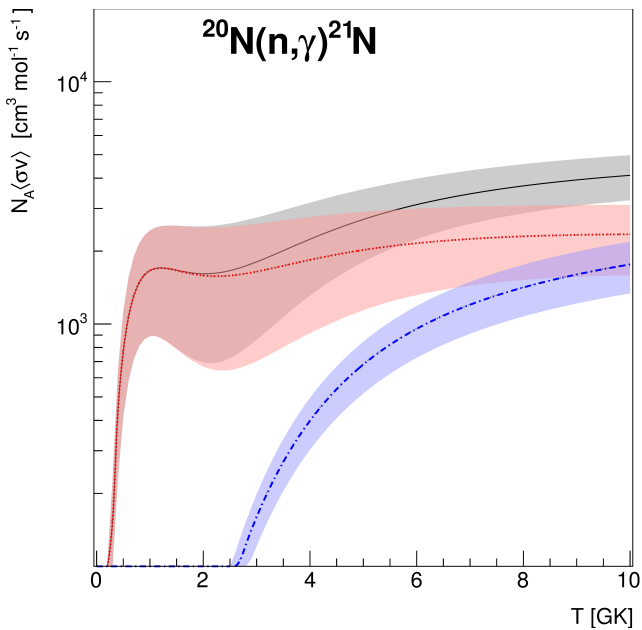}
  \caption{ 
      Color online. 
      Stellar reaction rate for $^{20}\mathrm{N}(\mathrm{n},\gamma)^{21}\mathrm{N}$ (black, solid line).
      The red, dotted line denotes contributions from the neutron capture of the ground state of $^{20}$N
      and the blue, dotted-dashed line denotes contributions from the neutron capture of the first and second excited state of $^{20}$N.
  }
  \label{fig:ReactionRate_20N-ng-21N}
\end{figure}

In Figure \ref{fig:ReactionRate19N-ng-20N}, the stellar reaction rate 
(compare Eq. \ref{eq:StellarReactionRate}) 
of $^{19}\mathrm{N}(\mathrm{n},\gamma)^{20}\mathrm{N}$ is plotted as a function of the temperature.
The red, dotted line denotes contributions from the ground state of $^{19}$N (generated with the randomization technique) and the blue, dotted-dashed line denotes contributions from the first excited state of $^{19}$N while considering the population probability (compare eqs. \ref{eq:Population1st19N5GK} and \ref{eq:StellarReactionRate}). 
The black, solid line denotes the stellar reaction rate
which is obtained by summing the ground-state and first excited-state contributions, 
while the black, dashed line denotes theoretical estimates \cite{Rauscher1994}.
It is clear that the excited state only plays a role for $T>3$\,GK.

At temperatures between 0.1 and 1.0\,GK, our measured data exceed the theoretical ones by up to a factor of three before the ratio drops below one and finally levels out at 0.1 of the theoretical predictions.

In Figure \ref{fig:ReactionRate_20N-ng-21N}, the stellar neutron capture reaction rate of $^{20}$N is shown in an analogous fashion to Figure \ref{fig:ReactionRate19N-ng-20N}.
Again, we note that due to the limited resolution of the gamma spectra, we could not distinguish between the first and second excited state of $^{20}$N. 
From Equation \ref{eq:StellarReactionRate}, one can see that this causes no major problem for the computation of the stellar reaction rate.
As the spins of the first and second excited states in $^{20}$N are equal \cite{Sohler2008}, only the difference in energy contributes to an additional uncertainty. 
We estimate this additional uncertainty to be 2\,\% of the stellar reaction rate, much less than the statistical uncertainty of $\ge$ 10\,\%. 
For lower temperatures, the contribution of the uncertainty due to the limited energy resolution is even lower, as the population probability decreases with decreasing temperature.
No theoretical predictions were available for this reaction \cite{Rauscher1994}.
The references given by Terasawa et al. in 2001 \cite{Terasawa_2001} 
do not include this reaction rate.
Therefore, we show the stellar reaction rate without any comparison. 

In Figure \ref{fig:ReactionRate_20N-ng-21N}, one can see that the stellar neutron capture reaction rate of $^{20}$N is of the same order of magnitude compared to the one of $^{19}$N.
For temperatures $T\le$0.8\,GK, the stellar neutron capture reaction rate of $^{20}$N is lower than the one of $^{19}$N. 
At a temperature of 5.5\,GK, the ratio of the stellar neutron capture reaction rates reaches its maximum where the rate of $^{20}$N is three times larger than the one of $^{19}$N. For higher temperatures, the ratio decreases to 2.5 at a temperature of 10\,GK.

\section{Network Calculations}
\label{sec:NetworkCalculations}

In order to illustrate the impact of our measured reaction rates on the final elemental abundance of an r-process calculation, we included our reaction rates in a reaction network code 
\cite{BradleyMeyer_NetworkCode_2012}.

As basic input, we downloaded the reaction network from the JINA Reaclib Database \cite{Cyburt_REACLIB_2010}. As several reactions on neutron-rich light nuclei were not included, we included the reaction rates of $^{14}$B(n,$\gamma$)$^{15}$B, $^{17}$C(n,$\gamma$)$^{18}$C, $^{18}$C(n,$\gamma$)$^{19}$C, $^{19}$C(n,$\gamma$)$^{20}$C, and $^{18}$C($\alpha$,n)$^{21}$O from Sasaqui et al \cite{Sasaqui_2005}. 
The remaining reaction rates in \cite{Sasaqui_2005} were of references predating our reference network.
In Ref. \cite{Sasaqui_2005}, Hauser-Feshbach models were used to obtain estimates on the reaction rates for light neutron-rich nuclei.
As the network calculation presented here is not supposed to explain the entire r-process nucleosynthesis, but only to give a rough impression of the impact of our measured reaction rates on the r-process, we omitted to derive new theoretical reaction rates.

Furthermore, we included the measured stellar reaction rate of $^{20}$N(n,$\gamma$)$^{21}$N (see previous section) and updated the rate of $^{19}$N(n,$\gamma$)$^{20}$N (referred to as ``our network'' in the following).
For comparison we created a second network (referred to as ``reference network'') which included the reaction rate of $^{19}$N(n,$\gamma$)$^{20}$N derived by Rauscher et al. \cite{Rauscher1994} but included our rate of $^{20}$N(n,$\gamma$)$^{21}$N as no previous reference for the latter reaction was available in the literature.

In order to simulate an r-process environment, we used the following trajectory for the evolution of density in time consisting of an exponential decay, reflecting the expanding medium of the supernova, 
and an additional slower decaying term, reflecting the neutrino wind \cite{Meyer_2002_trajectory}
\begin{equation}
 \rho(t) = \rho_0 \exp(-t/\tau_0) + \frac { \rho_1 } { (1+t/\tau_1)^2} \, ,
 \label{eq:DensityEvolution}
\end{equation}
where 
$\rho_0 = 1.7995 \cdot 10^6$\,g/cm$^3$ is the initial density (at t = 0\,s) of the exponential term,
$\rho_1 = 540$\,g/cm$^3$ is the constant density of the second term,
$\tau_0 = 0.0051$\,s is the decay constant of the exponential term.

The temperature $T$ is parameterized as
\begin{equation}
 T(t) \propto \rho(t)^{1/3} \, ,
 \label{eq:TemperatureEvolution}
\end{equation}
with $T(t = 0\,\mathrm{s}) = 9.0$\,GK.
These input parameters were chosen to match Terasawa et al. \cite{Terasawa_2001}.
The initial neutron to proton ratio was set to 65/35 (which corresponds to set 2 in table 4 of Sasaqui et al. \cite{Sasaqui_2005}).

\begin{figure}[t!]
  \includegraphics[width = 0.47\textwidth]{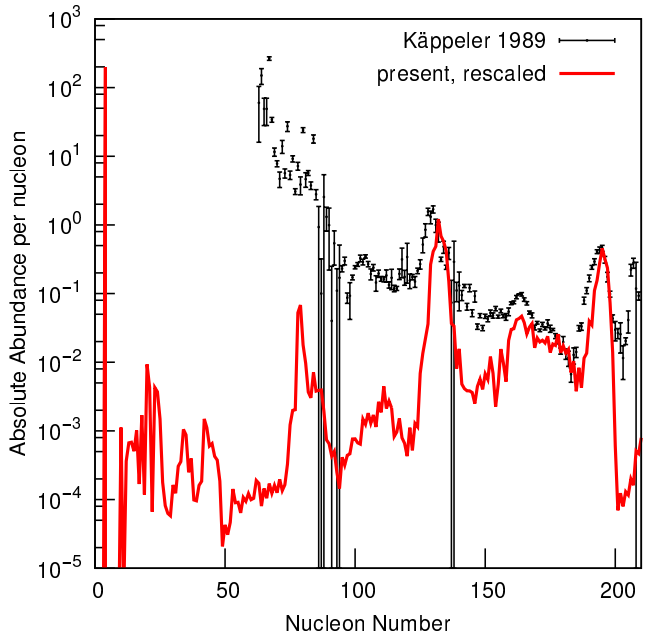}
  \caption{ Color online. 
    Final abundances as a function of the nucleon number derived from our network calculation with our measured reaction rates (red, solid line) and solar r-process abundance from reference (black symbols) \cite{Kaeppeler1989}. For details, see text.
  }
  \label{fig:Network_AbsoluteAbundance}
\end{figure}

\begin{figure}[t]
  \includegraphics[width = 0.47\textwidth]{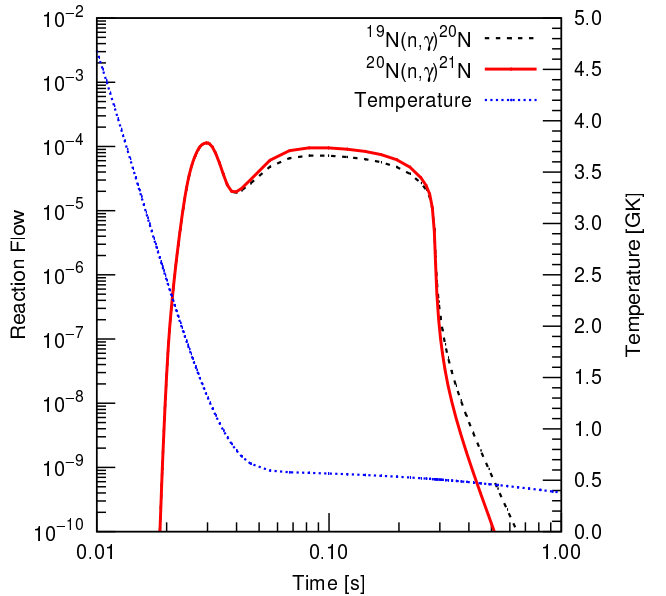}
  \caption{ Color online. 
    Reaction flow as a function of time (and temperature) during our network calculation for $^{19}$N(n,$\gamma$)$^{20}$N (black, dashed line) and $^{20}$N(n,$\gamma$)$^{21}$N (red, solid line). The temperature is plotted in blue, dotted.
  }
  \label{fig:Network_ReactionFlow}
\end{figure}

Then, the network was run to calculate the reaction flow until $10^6$\,s after the supernova exploded.
In Figure \ref{fig:Network_AbsoluteAbundance}, we extracted the abundance as a function of nucleon number at that time (red line) where the three main r-process peaks are visible at A = 79, A = 130, and A = 195.
For comparison we added the solar abundances of r-process elements as derived by Käppeler et al. 1989 \cite{Kaeppeler1989} (black symbols) and rescaled our data to match the height of the peak at A = 195.

Our efforts on network calculations are not to understand the entire r-process in detail but to estimate the impact of our measured reaction rates on the r-process abundances.
Therefore, we use the abundances derived with our network for comparison with abundances computed when using the reference network (see previous passage).

We find a decrease of 10\,\% for fluorine ($A = 19$),
but the abundances of nuclei with higher mass are unaffected by our data.

In Figure \ref{fig:Network_ReactionFlow}, we show the reaction flow which is the contribution per nucleon per second to the abundance of one of the product species in the reaction. 
One can see that the reactions under study become important (e.g. with a reaction flow $\geq 10^{-5}$) at times within $t \in [0.025, 0.35]$\,s which correspond to temperatures $T \in [2.07, 0.60]$\,GK.
Thus, we measured the reaction rate exactly in the important temperature range.
Corrections due to neutron capture of excited states of the target nucleus play no role in this temperature range (Figures \ref{fig:ReactionRate19N-ng-20N} and \ref{fig:ReactionRate_20N-ng-21N}).

In order to gauge the possible impact of an increase in the reaction rate due to neutron capture into excited states of the product nucleus, which is not constrained by the present data (see sec. \ref{sec:AstrophysicalReactionRate}), the network calculations have been repeated with tenfold increased (n,$\gamma$) reaction rates. The resulting fluorine abundance is 50\% lower than in the reference case, and again no measurable impact is seen on any other nucleus.

\section{Summary and Outlook}
\label{sec:Summary}

\noindent 
\textbf{Summary:}
We measured the Coulomb dissociation cross section of $^{20,21}$N
and discriminated between transitions into the ground state and the first excited state of the outgoing nuclei.
In the case of $^{21}$N, we could not separate the first from the second excited state due to the limited resolution of our gamma calorimeter. Therefore, we presented only transitions into any excited state of the outgoing $^{20}$N ions.

Furthermore, we calculated the photo absorption cross sections via the virtual photon theory and the neutron capture cross section by using the principle of detailed balance for each individual reaction, discriminating for the excitation level of the outgoing particle.
In these cases, the ground state contributions had such a low statistical uncertainty that only an upper limit could be presented.

Moreover, the thermonuclear and the stellar reaction rates were computed even for ground state transitions due to the use of a Monte Carlo method.
Additionally, the reaction rates were compared to theoretical predictions.

Finally, network calculations were performed to estimate the impact of our measured reaction rates on the possible r-process scenario of a supernova with a neutrino driven wind.

A decrease of 10\,\% in the fluorine ($A = 19$) abundance was found
relative to the abundances when using the reference rates \cite{Rauscher1994}. The abundances of nuclei with higher mass were unaffected.

\noindent 
\textbf{Outlook:}
In the future, our measured reaction rates may be implemented in reaction networks which include even more light neutron-rich nuclei. Other possible r-process scenarios could be used to study the impact of these reaction rates on the final r-process elemental abundance in more detail.

A repetition of our measurements with more statistics could reduce the statistical uncertainties already in the excitation energy spectra and the photo absorption and neutron capture cross sections.

Additionally, the low resolution of our gamma spectra prevented the separation of the first from the second excited state of outgoing $^{20}$N.
Thus, a repetition with a 
gamma spectrometer with higher granularity and intrinsic energy resolution
would improve the separation on the one hand and the resolution of the excitation energy spectra on the other hand.

Both experimental improvements will be provided in the future FAIR (Facility for Antiproton and Ion Research) which is presently under construction at GSI in Darmstadt, Germany \cite{Reifarth13-NPA6}.

\begin{acknowledgments}

This work was supported in part by 
GSI (F\&E, DR-ZUBE),
BMBF (06DR134I, 05P09CRFN5),
the Nuclear Astrophysics Virtual Institute (NAVI, HGF VH-VI-417), 
the Helmholtz Association Detector Technology and Systems Platform,
Spanish Research funding agency under projects FPA2012-32443, FPA2013-41267-P, and FPA2013-47831-C2-1-P,
the Swedish Research Council,
BMBF(05P12RDFN8 and 05P15RDFN1), 
HIC for FAIR and the TU Darmstadt-GSI cooperation contract,
the Portuguese FCT project PTDC/FIS/103902/2008,
UK STFC under grants
ST/E500651/1 and ST/F011989/1, 
U.S. NSF Grant No. 1415656, 
and 
U.S. DOE grant No. DE-FG02- 08ER41533.

\end{acknowledgments}

\end{document}